\documentstyle[preprint,aps,prb,epsf,amsfonts]{revtex}

\newcommand{\eps}{\varepsilon}
\newcommand{\Bcr}{B_{\rm cr}}
\begin{document}

\draft
\title{Universal properties for linelike melting of the vortex 
lattice} 
\author{A. E. Koshelev and H. Nordborg } 
\address{Materials Science Division, Argonne National Laboratory, 
Argonne, Illinois 60469} 
\date{\today} 
\maketitle

\tighten
\begin{abstract}

Using numerical results obtained within two models describing vortex 
matter (interacting elastic lines (Bose model) and uniformly 
frustrated XY-model) we establish universal properties of the melting 
transition within the linelike regime.  These properties, which are 
captured correctly by both models, include the scaling of the melting 
temperature with anisotropy and magnetic field, the effective line 
tension of vortices in the liquid regime, the latent heat, the entropy 
jump per entanglement length, and relative jump of Josephson energy at 
the transition as compared to the latent heat.  The universal 
properties can serve as experimental fingerprints of the linelike 
regime of melting.  Comparison of the models allows us to establish 
boundaries of the linelike regime in temperature and magnetic field.
\end{abstract}

\pacs{74.20.De,74.60.Ec,74.60.Ge}
\narrowtext


It is now well established that the vortex lattice in a clean 
superconductor melts in a first-order transition due to thermal 
fluctuations.  The melting transition has been observed in a number of 
experiments using different techniques, including transport 
\cite{Safar-etal:PRL92,Kwok-etal:92,Charalambous-etal:PRB92} and 
direct measurements of the magnetization jump 
\cite{Zeldov-etal:NAT95,Liang-etal:PRL96,Welp-etal:PRL96} and the 
latent heat \cite{Schilling-etal:NAT96,Roulin-etal:SCI96} associated 
with the melting transition.  

Melting of the vortex lattice in real superconductors is a complicated 
phenomenon.  Several different regimes exist depending on the 
parameters of the superconductors and the magnetic field.  The 
simplest and most natural picture of the melting transition in 
three-dimensional superconductors is that a low temperature crystal of 
straight vortex lines transforms into a liquid of entangled 
lines.\cite{Nelson:PRB89} In real layered superconductors this picture 
does not work either at too high or too low fields.  At high fields, 
the vortex lattice is more adequatly described as a system of weakly 
interacting two-dimensional lattices.  Melting in this situation 
occurs in a pointlike way and is accompanied by evaporation of vortex 
lines.\cite{GlazKosh:PRB91} At low fields, on the other hand, the 
melting temperature approaches the fluctuation region and vortex 
degrees of freedom start to interact with other superconducting 
degrees of freedom (spin wave excitations and vortex loops).  The 
contribution of these degrees of freedom to the latent heat of the 
melting starts to increase abruptly as the temperature aproaches 
$T_{c}$ possibly leading to a significant change in the nature of the 
melting transition.\cite{Tesanovic:PRB95,Nguyen-Sudbo:PRB98a}

As a quantitative theory of vortex lattice melting has yet to be 
developed, a large amount of work has been invested into numerical 
simulations of the vortex system.  A complete picture can be worked 
out by direct simulation of the Ginzburg-Landau functional, but this 
would require enormous computational effort.
Rather a number of simplified models have been used, such as 
interacting elastic lines (or the Bose model) 
\cite{Ryu-etal:PRL92,Ryu-Stroud:PRB96,Magro-Ceperley:PRB93,Magro-Ceperley:PRL94,%
Nordborg-Blatter:PRL97,Nordborg-Blatter:PRB98}, the frustrated XY-model 
\cite{Hetzel-Sudbo-Huse:PRL92,Li-Teitel,Chen-Teitel:PRB97,%
Hu-etal:PRL97,Nguyen-Sudbo:PRB98a,Nguyen-Sudbo:PRB98b}, 
the Lattice London Model (or 
lines and loops) \cite{Cavalcanti,Carneiro:1995,Chen-Teitel:PRL95}, 
and the Lowest Landau 
Level approximation.\cite{Sasik-Stroud:PRL95,Hu-MacDonald:PRB97} 
Being approximations, all these models have their limited regime of 
applicability where they can be expected to make reliable predictions 
for the melting transition and for the properties of different vortex 
phases.  Unfortunately, it is difficult even to make reliable 
analytical predictions for the range of applicability of a specific 
model.  In this work we therefore use a more pragmatic approach of 
comparing the numerical results for two rather different models: the 
Bose model and the XY-model.  We find that the results obtained from 
these models agree over a wide range of fields and temperatures, where 
the melting transition can be described as \emph{linelike}.  We show 
that in the linelike regime the transition is characterized by a set 
of \emph{universal properties}, which can serve as fingerprints of 
this regime.  Comparison of the models also allows us to establish 
boundaries of the linelike regime.

\section{Description and comparison of the models}

The phenomenology of superconductors is based on the Ginzburg-Landau 
model, which expresses the energy of superconductor via the magnetic 
induction, $\mathbf{B}$, the modulus of the order parameter, $|\Psi|$, 
and its phase, $\phi = \arg \Psi$.  The model is completely defined by 
the London penetration depth $\lambda$, the coherence length $\xi$, 
and the anisotropy factor $\gamma$.  Since the high-$T_c$ materials 
are strongly type II superconductors, i.e., $\kappa = \lambda / \xi 
\gg 1$, the modulus of the order parameter is suppressed only in the 
vicinity of vortex cores.  This allows us to describe superconductors 
within the \emph{London approximation} which neglects fluctuations in 
the modulus of the order parameter everywhere except in the cores of 
the vortices.  The approximation breaks down in the vicinity of the 
upper critical field $H_{c2}$ and in the fluctuation region near the 
transition temperature $T_{c}$.  The London approximation is used by 
both models studied in this paper.  The frustrated XY-model then makes 
two further approximations: It neglects the fluctuations in the 
magnetic induction, usually called the \emph{frozen-field} or 
\emph{infinite-lambda} approximation, and introduces a lattice on 
which the phase of the order parameter is defined.  The model desribes 
layered superconductors only in terms of the phase distribution $\phi({\bf 
n})$, defined on a three-dimensional grid ${\bf n}=(n_{x},n_{y}, 
n_{z})$.  The energy functional of this model is given by (see, e.g., 
Ref.~\onlinecite{Koshelev:PRB97}).  
\begin{equation}\ {\cal 
F}[\phi({\bf n})] =  \sum_{\mathbf{n}} \left\{ J \sum_{\alpha = x,y} 
V\left[ \phi({\bf n} + {\bf d}_\alpha) - \phi({\bf n}) - a_\alpha({\bf 
n})\right] - \frac{J}{\gamma^2} \cos \left[ \phi({\bf n} + {\bf d}_z) 
- \phi({\bf n})\right] \right\},
\label{xyaction}
\end{equation}
The energy scale of the system is the phase stiffness, $J$, definied 
as
\begin{equation}
J = \frac{s\Phi_0^2}{\pi (4\pi\lambda_{ab})^2} \equiv 
\frac{s\eps_0}{\pi},
\end{equation}
where $s$ is the layer spacing and we have introduced the scale for the 
vortex line energy $\eps_0 = (\Phi_0/4\pi\lambda_{ab})^2$ with 
$\Phi_0$ being the flux quantum and $\lambda_{ab}$ is the penetration 
depth for supercurrents flowing in the $ab$-planes.  The anisotropy is 
given by $\gamma = \lambda_c/\lambda_{ab}$, where $\lambda_c$ is the 
penetration depth for supercurrents flowing the direction along the 
$c$-axis.  The phase $\phi({\bf n})$ is defined on a cubic lattice 
with the lattice spacing $s$, the dimensionless vector potential is 
defined as ${\bf a} = (0, 2\pi f n_{x} / \Phi_0, 0)$, with $f$ being 
the fraction of the lattice cells filled by vortices, and ${\bf 
d}_\alpha$ are unit vectors.  The phase interaction $V(\phi)$ is a 
$2\pi$-periodic function with the Taylor expansion $V(\phi) - V(0) 
\approx \phi^2/2$ as $\phi\rightarrow 0$.  An obvious choice is to use 
$V(\phi) = -\cos(\phi)$, but this produces large barriers for the 
motion of the vortices between the lattice cells.  A natural 
improvement is therefore to choose a function which minimizes these barriers, as 
has been done in Ref.~\onlinecite{Koshelev:PRB97}.

The Bose model, on the other hand, describes the vortex system
entirely in terms of the vortex degrees of freedom.  A vortex is 
treated as an elastic string, interacting with the other lines through
a \emph{screened} Coulomb potential represented by the modified Bessel
function ${\rm K}_0(R/\lambda_{ab})$.  The free energy functional is
then given 
by \cite{Nordborg-Blatter:PRB98}
\begin{equation}
{\cal F}[{\bf R}_i(z)] = \int_0^{L_z} dz \left\{ \sum_i
\frac {\varepsilon_l} {2} \left( \frac{d{\bf R}_i}{d z} \right)^2 
+ \sum_{i \neq j} \varepsilon_0 {\rm K}_0 \left( \frac{R_{ij}}
{\lambda_{ab}}\right)
\right\},
\label{boseaction}
\end{equation}
with $L_z$ is the thickness of the sample, ${\bf R}_i(z)$ is a 
two-dimensional vector describing the position of the vortex, and 
$\eps_l$ is elasticity of a vortex line.  This single-line elasticity 
is an effective quantity, which we choose to reproduce the relevant 
tilt modulus of the vortex lattice.  Two effects complicate the
relation between this parameter and the parameters for
superconductors. First, it is well known that the tilt energy of
vortex line is nonlocal and the effective line stiffness for deformations
with wave vector $k_{z}$ is proportinal to $\ln(\gamma/k_{z}\xi)$.  The
$k_{z}$ relevant for melting are given by $k_{z}\sim \gamma/a_{0}$.
This leads to the estimate \cite{Nordborg-Blatter:PRB98}
\begin{equation}
\eps_l \approx \frac{\eps_0}{\gamma^2} \ln \left( 
\frac{a_0}{2\sqrt{\pi}\xi} \right),
\label{logdef}
\end{equation} 
where $a_0$ is the lattice spacing of a triangular lattice, i.e., 
$\Phi_0/B = \sqrt{3}a_0^2/2$. The second effect becomes important in 
the temperature range close to the fluctuation region where phase and
vortex fluctuations start to suppress the Josephson coupling. This 
suppression enhances the anisotropy and reduces the effective line
tension of vortices $\eps_l$ as compared to the estimate
(\ref{logdef}). In the following we introduce the effective anisotropy
parameter $\eps$ of the Bose model as $\eps^2 \equiv \eps_l/\eps_0$,
bearing in mind that $\eps \sim 1/\gamma$, with the main difference
coming from the logarithm in Eq.\ (\ref{logdef}).

Both models have their limitations in describing real superconductors.  
The Bose model describes a continuous anisotropic system whereas the 
XY-model takes the layered structure of the superconductor into 
account.  Thus one would not expect the Bose model to be applicable 
when the typical wavelength in the $z$-direction becomes comparable to 
the layer spacing.  As will be shown below, the correlation length in 
the $z$-direction for the Bose model is $l_z = \alpha a_0 / \gamma$, 
where $\alpha \approx 6$ is a numerical constant.  \cite{Teitel:98} 
Therefore, we expect the Bose model to be invalid when $a_0 < \gamma s 
/ \alpha$ or for $B \gtrsim \alpha^2\Bcr$.  Numerical results show 
that line melting works at least for fields as large as 
$10\Bcr$.\cite{Koshelev:PRB97} For larger fields, the Josephson 
coupling between the layers is strongly suppressed by thermal 
fluctuations, invalidating the use of a simple line model.
  
Both models neglect the electromagnetic coupling between the 
superconducting layers.  This coupling is relevant for very 
anisotropic materials when $\gamma s > a_0, 
\lambda$.\cite{Blatter-etal:PRB96} The XY-model further neglects the 
screening of the vortex interaction due to the magnetic field, which 
becomes relevant close to $T_c$ when $a_0 \gtrsim \lambda(T)$, and 
therefore does not describe the reentrant behavior of the melting line 
which is expected for low fields.

When the melting temperature approaches the fluctuation region 
thermally activated vortex loops start to influence the 
thermodynamic properties of superconductor strongly.  The Bose model only 
contains the field induced vortices as degrees of freedom and neglects 
this effect.  The XY-model allows for both Gaussian phase 
fluctuations and thermally induced vortex loops in addition to the 
field induced vortices.  The extra degrees of freedom give a large 
contribution to the specific heat of the system: Typical numerical 
data from the XY-model show a specific heat which grows with 
temperature and has a broad maximum at $T_c$.  On top of this specific 
heat there is a small sharp peak due to the first-order melting 
transition.\cite{Hu-etal:PRL97,Nguyen-Sudbo:PRB98a} These results 
agree well with experimental 
observations.\cite{Schilling-etal:NAT96,Roulin-etal:SCI96} The 
important issue is to understand how the field-induced vortices couple 
to the other fluctuations. Outside the fluctuation region, the coupling 
only leads to a weak renormalization of vortex interaction and the effective
line energy, as is apparent from a number of observations:  To 
begin with, the shape of the melting line is well 
reproduced by the Lindemann criterion, which only includes field 
induced vortices with unrenormalized interactions 
(see, e.g., Refs.~\onlinecite{Houghton-etal:PRB89} 
and \onlinecite{Kwok-etal:92}).  Furthermore, it has recently been 
shown that the field and temperature dependence of the discontinuity at 
the vortex lattice melting transition also can be estimated using a 
simple line model and the mean-field temperature dependence of 
$\eps_0(T)$.\cite{Dodgson-etal:PRL98}. On the other hand, recent simulations
have shown that if melting takes place in the vicinity of the 
fluctuation region, then a liquid phase immediatly above the transition 
is not simply a liquid of interacting elastic lines, as the Bose model 
assumes, but rather an infinite interconnected cluster, in which 
individual lines loose their identities.\cite{Nguyen-Sudbo:PRB98a} 
Therefore it is natural to test the adequacy of the line picture and 
establish boundaries of its applicability by direct comparison of 
the numerical results of the Bose model with the more microscopic 
XY-model.

\section{Universal properties. Comparison of numerical results}

\begin{figure}
\centerline{\epsfxsize= 8.5cm\epsfbox{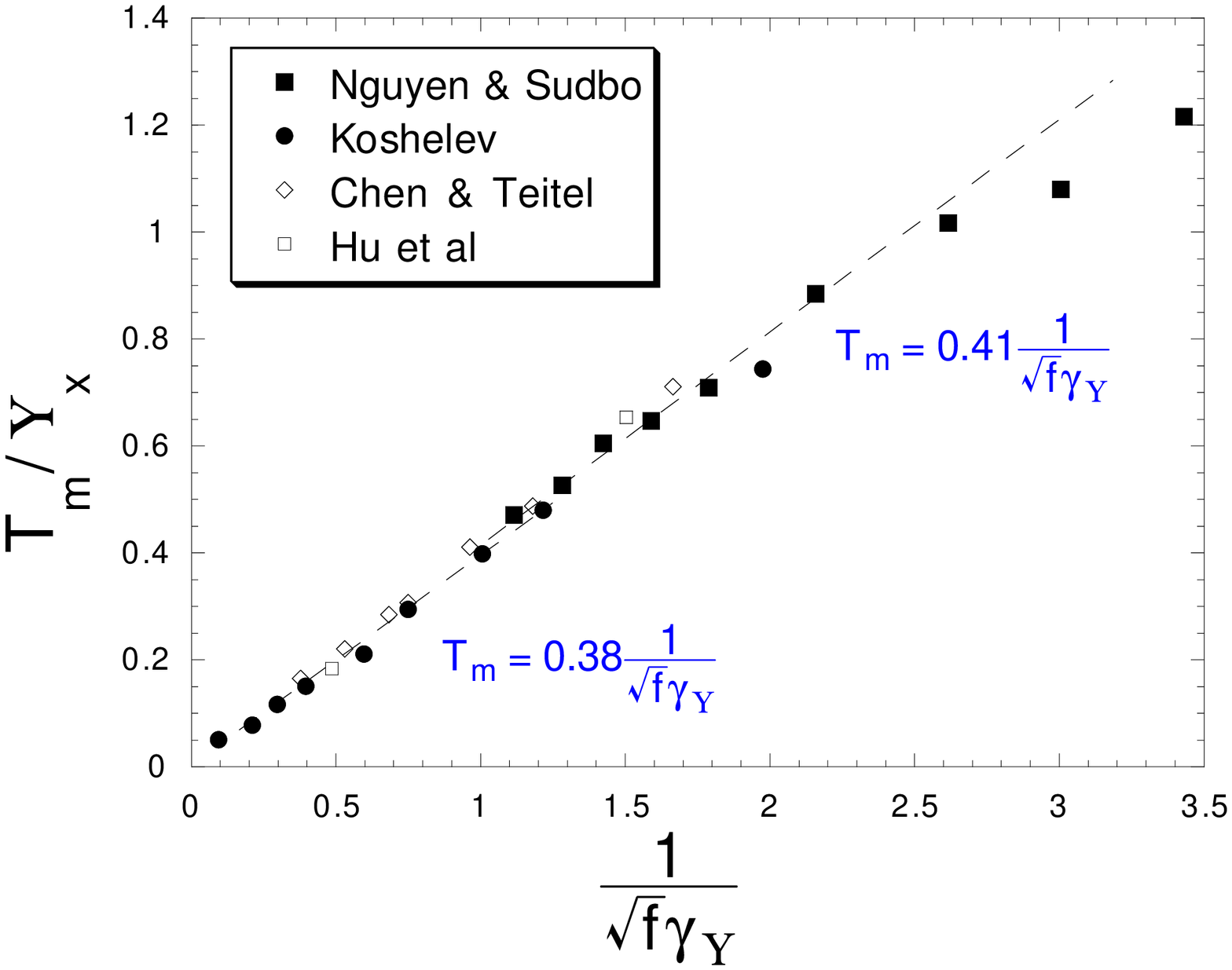}}
\caption{Scaling plot of melting temperature in the line melting 
regime for XY models with different anisotropy factors.  Data are 
taken from 
Refs.~\protect\onlinecite{Chen-Teitel:PRB97,Koshelev:PRB97,Hu-etal:PRL97,Nguyen-Sudbo:PRB98b}.
Linear fits are made for data from Refs.~\protect\onlinecite{Koshelev:PRB97,Nguyen-Sudbo:PRB98b}.}
\label{meltscalXY}
\end{figure}

Important conclusions can be drawn already from the relevant scales 
for energy, length, and magnetic field in the expressions 
(\ref{xyaction}) and (\ref{boseaction}).  For the XY-model, the 
relevant scale for the magnetic field is the crossover field $\Bcr = 
\Phi_0/(\gamma s)^2$ and the relevant energy scale is the phase 
stiffness $J$.\cite{Koshelev:PRB97} Thus, melting lines for systems 
with different anisotropy factors should collapse on to one line as long 
as the field is measured in units of $\Bcr$ and temperature is 
measured in units of $J$, as has indeed been demonstrated 
experimentally.\cite{Sasagawa-etal:PRL98,Welp-scaling} In the region 
of small fields $B\lesssim \Bcr$ and beyond the fluctuation region $T 
\lesssim J$ the lattice is expected to melt in a linelike fashion, 
i.e., the lines are expected to retain their identity above the 
transition and the Josephson coupling between the layers is only 
weakly suppressed.  In this regime one can expand the Josephson 
coupling energy with respect to interlayer phase difference and make 
a continuous approximation, which leads to an even simpler scaling 
property: The melting temperature should scale as
\begin{equation}
T_{m} = A_{m} J\sqrt{\Bcr/B}.
\label{meltlineXY}
\end{equation}
Simulations show that this scaling indeed works very well for fields 
$\Bcr \lesssim B \lesssim 10\Bcr$ with $A_{m}\approx 0.33$ even though 
a noticable suppression of the Josephson coupling is observed near 
$T_{m}$ (down to 64\% of the bare coupling).
\cite{Koshelev:PRB97} The scaling form (\ref{meltlineXY}) assumes, in 
fact, that only vortex degrees of freedom participate in the melting 
transition and regular phase fluctuations play negligible role at 
$T_{m}$.  This approximation breaks down when the melting temperature 
approaches the fluctuation region, which corresponds to $T_{m}\gtrsim 
J$ or $B\lesssim \Bcr$.  Regular phase fluctuations and thermally 
activated vortex loops suppress both the in-plane phase stiffness and 
the Josephson interlayer coupling.  This leads to a renormalization of 
vortex interactions and their tilt stiffness.  Quantitatively these 
effects can be characterized by the {\em helicity moduli} 
$\Upsilon_{x}$ and $\Upsilon_{z}$ in the Meissner state.  A natural 
generalization of the scaling form (\ref{meltlineXY}), which takes 
into account this renormalization, can be obtained by the replacements 
$J\rightarrow \Upsilon_{x}$ and $\gamma \rightarrow \gamma_{{\tiny 
Y}}$ with $\gamma_{{\tiny Y}}^{2}=\Upsilon_{z}/\Upsilon_{x}$ being the 
anisotropy of the helicity moduli.  This leads to a generalized 
scaling relations for melting temperature
\begin{equation}
T_{m} = \tilde{A}_{m} \Upsilon_{x}\sqrt{\Phi_{0}/{B(\gamma_{{\tiny Y}} s)^{2}}}
\label{genscalXY}
\end{equation}
For most of the phase diagram melting occurs in the region where 
regular phase fluctuations are weak and one can use formulas for 
$\Upsilon_{x,z}$ with small fluctuation corrections (see Appendix)
\begin{eqnarray}
\Upsilon_{x}&\approx &J\left[1-\frac{V^{(4)}T}{4J}\left(1-\frac{\ln(32 
\gamma^{2})-1}{2\pi\gamma^{2}}\right)\right], \\
\Upsilon_{z}&\approx & {\frac{J}{\gamma^{2}}}\left[1-\frac{T(\ln(32 
\gamma^{2})-1)}{4\pi J}\right]
\label{corrections}
\end{eqnarray}
with $V^{(4)}=-\partial^{4} V(\theta)/\partial \theta^{4}$ at 
$\theta=0$ and $V(\theta)$ is the phase interaction function from 
Eq.~(\ref{xyaction}).  To check the scaling relation (\ref{genscalXY}) 
we used all availible data on the melting transition for XY models with 
different anisotropies and filling factors.  
\cite{Chen-Teitel:PRB97,Koshelev:PRB97,Hu-etal:PRL97,%
Nguyen-Sudbo:PRB98b} 
Fig.~\ref{meltscalXY} shows the dependence of $T_{m}/\Upsilon_{x}$ vs 
$1/\sqrt{f} \gamma_{{\tiny Y}}$, which according to Eq.~(\ref{genscalXY})
should be a straight line.  We found that in a suprisingly wide range
of fields and anisotropies, including a substantial part of the fluctuation
region, the scaling (\ref{genscalXY}) works very well with 
$\tilde{A}_{m}\approx 0.4$. At low fields the generalized 
scaling relation extends down to $B\approx 0.1 B_{cr}$.

Near the transition point $T_{c}=\alpha_{c}J$ the helicity moduli vanish 
according to the XY-scaling laws
\begin{eqnarray}
\Upsilon_{x}&\approx &J\left(\alpha_{c}-\frac{T}{J}\right)^{2\nu}, \\
\Upsilon_{z}&\approx & {\frac{J}{\gamma^{2}}}\left(\alpha_{c}-\frac{T}{J}\right)^{2\nu},
\label{XYscaling}
\end{eqnarray}
with $\nu\approx 1/3$.  Note that above representations include the
possibility of a temperature dependent $J$ near the mean field 
transition temperature $T_{c0}$, $J=J_{0}(T_{c0}-T)/T_{c0}$.  
Assuming that the scaling law (\ref{genscalXY}) can be extended to 
the fluctuation region we conclude that in the vicinity of the transition 
temperature $T_{c}$ the melting field should scale as $B_{m}\propto 
(T_{c}-T)^{4/3}$.  This scaling indeed describes the behavior of 
experimental melting line better than the ``mean field'' scaling 
$B_{m}\propto (T_{c0}-T)^{2}$, which follows from scaling relation 
(\ref{meltlineXY}) (see, 
Refs.\onlinecite{Kwok-etal:92,Liang-etal:PRL96,Welp-etal:PRL96}).

The Bose model is able to describe the melting transition for 
arbitrary ratio $a_{0}/\lambda$.  There is no general scaling relation 
which holds for arbitrary values of this ratio.  However, in the 
region of the high vortex density $a_{0}\ll \lambda$, where the 
interaction can be approximated with a logarithm, the free energy only 
depends on the dimensionless parameter $\Lambda = 
T/a_0\sqrt{2\eps_l\eps_0}$.\cite{Nordborg-Blatter:PRB98} In 
particular, the melting takes place at $\Lambda_m= 0.0622$ 
\cite{Nordborg-Blatter:PRB98} giving the following scaling relation 
for the melting temperature.
\begin{equation}
T_m \approx 0.088 \eps\eps_0 a_0.
\label{BoseT}
\end{equation}
This scaling form corresponds to the low field scaling of the XY-model 
given by Eqs.~(\ref{meltlineXY}) and (\ref{genscalXY}). Rewriting
Eq.~(\ref{meltlineXY}) in terms of $a_0$ we obtain 
\begin{equation}
T_m \approx \frac{0.33}{\pi}\sqrt{\frac{\sqrt{3}}{2}} \frac{\eps_0 
a_0}{\gamma} \approx 0.098 \eps_0 a_0 /\gamma.
\label{XYT}
\end{equation}
Thus, the two models do not only agree on the shape of the melting 
line, but they also agree quantitatively on the position of the 
melting transition. The result is consistent with the assumption 
that, according to Eq.~(\ref{logdef}), we have $\eps > 1 / \gamma$ and we 
find
\begin{equation}
\eps \approx 1.11 / \gamma.
\label{epsdef1}
\end{equation}

For finite range interactions the melting temperature is lowered; 
simulations with $\lambda = 1.06 a_0$ give $T_m \approx 0.084 
\eps\eps_0 a_0$, i.e., the melting temperature is lowered by 
approximately $5\%.$\cite{Nordborg-Blatter:PRB98}

\begin{figure}
\centerline{\epsfxsize= 8.5cm\epsfbox{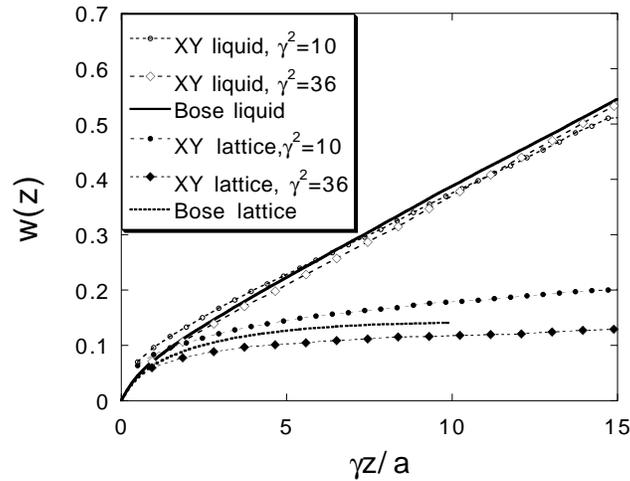}}
\caption{Comparison of  line wandering in the crystal and liquid states near the melting 
point for the Bose model and the frustrated XY-model. The two models show very similar 
behavior both with regards to the asymptotic and the initial slope in the liquid. 
Likewise, the saturation of the line wandering in the crystal is captured correctly
by both models.}
\label{fig1}
\end{figure}

We now compare properties of the line liquid near the transition.  Both 
experiments and simulations show the vortex liquid to be heavily 
entangled so that the phase coherence in the direction parallel to the 
applied field is destroyed by melting.  The entanglement can be 
quantified in terms of the transverse wandering of the vortex lines,
\begin{equation}
w(z) = \left\langle \left[ {\bf R}(z) - {\bf R}(0) 
\right]^2 \right\rangle/a_{0}^{2}.
\end{equation}
It is easy to show that we have 
\begin{equation}
w(z) = \frac{2T}{\eps_l a_0} \frac{z}{a_0},
\label{LinWan}
\end{equation}
for non-interacting lines. Since interactions between the lines are
unimportant for short distances, this is also the result for small $z$
in the Bose model. For larger $z$, the wandering is suppressed due to
the interaction between the vortices but we still expect $w(z) \propto
z\gamma$. The numerical result is
\begin{equation}
w(z) = 0.27 \frac{2T}{\eps_l a_0} \frac{z}{a_0},
\end{equation}
implying that the line stiffness is increased by a factor $3.7$ due 
to the interaction with other lines in the liquid. For the XY-model 
as well as for real superconductors the linear dependence 
(\ref{LinWan}) is only approximate due to nonlocality of the line 
tension. In Fig.~\ref{fig1} 
we show $w(z)$ for two different values of $\gamma$ in the XY-model, 
just above the melting transition. We also show $w(z)$ from the Bose 
model, where we have chosen $\eps$ to give the best fit. This yields,
\begin{equation}
w(z) \approx 1.46 z / \eps a_0 \Rightarrow \eps \approx 1.46 / \gamma,
\end{equation}
which again is in reasonable agreement with with Eq.~(\ref{logdef}).  
For comparison the plot also shows $w(z)$ in the lattice just below 
the melting transition.  One can see that line wandering agrees very 
well for the two models also in the lattice state.  In particular an, 
initial growth of $w(z)$ is similar to the liquid state and it starts to 
saturate at approximately the same length scale along $z$-direction.  
For the Bose model no entanglement is observed in crystal state and line 
wandering saturates at $w(z) \approx 0.15$. For the XY-model $w(z)$ 
continues to grow due to rare reconnection events, with slope much smaller 
than in the liquid state. This small slope increases with decreasing
anisotropy.

It is convenient to have a single number characterizing the line 
wandering in the vortex liquid.  A suitable quantity is the 
entanglement length $l_e$, which is defined as a length at which 
average line displacement becomes equal to the radius of the Bravais cell,
\begin{equation}
w(l_e) = \frac{\Phi_0}{B \pi a_0^2 } \approx 0.276. 
\end{equation}
From Fig.~\ref{fig1} we find an entanglement length $l_e \approx 6.6 
a_0 / \gamma$, implying that the vortex system can become fully 
entangled already in a thin sample. 

We now turn to the thermodynamic properties of the melting transition obtained 
from simulations.  A main characteristic of the melting transition is 
the latent heat or entropy jump.  As is now generally accepted, it 
is important to take the internal temperature dependence of the free 
energy functional ${\cal F}$ into account when computing the energy of 
a superconductor using Ginzburg-Landau 
theory.\cite{Hu-MacDonald:PRB97,Dodgson-etal:PRL98,Koshelev:PRB97} As 
a result, the energy is not just the average of ${\cal F}$ but is 
rather given by
\begin{equation}
E = \langle {\cal F} \rangle - T \left\langle \frac{\partial {\cal 
F}}{\partial T} \right\rangle,
\end{equation}
where $\langle\dots\rangle$ indicates the Monte Carlo average. In the 
XY-model, the temperature dependence of ${\cal F}$ is only due to the 
temperature of the energy scale $J = s\eps_0/\pi$. The same is true 
for the Bose model if we assume a logarithmic interaction between the 
vortices, i.e., when $\lambda > a_0$. We then obtain the simple result
\begin{equation}
E = \left(1- \frac{T}{\eps_0} \frac{\partial \eps_0}{\partial 
T}\right) \langle {\cal F} \rangle.
\end{equation}
This equation, together with numerical data for the discontinuity in 
$\langle {\cal F} \rangle$, produces results which are consistent 
with experiments.\cite{Dodgson-etal:PRL98} 

It should be remembered, however, that the statistical weight of a 
particular vortex configuration is still given by $\cal F$. It 
therefore makes sense to study the \emph{configurational} energy $E_c
= \langle {\cal F} \rangle$ and the corresponding entropy, $S_c = E_c
/ T = \langle {\cal F} \rangle/T.$ It is this configurational entropy 
which we expect to be proportional to the number of degrees of 
freedom of the vortex system. In order to be able to compare the results 
from the two models, we only consider the energy per vortex and length, 
\begin{equation}
e_c = \frac{E_c}{NL_z} = \frac{1}{NL_z} \langle {\cal F} \rangle. 
\end{equation}
The Bose model gives\cite{Nordborg-Blatter:PRB98}
\begin{equation}
\Delta e_c \approx 0.013 \eps_0
\end{equation}
and the XY-model typically produces a jump\cite{Koshelev:PRB97}
\begin{equation}
\Delta e_c \approx \ 0.016 \eps_0.
\end{equation}
It is interesting to note that the jump in the configurational 
entropy per entanglement length at the transition is 
\begin{equation}
\Delta s_{c} l_e = \Delta e_c l_e / T_m \approx (0.9-1.1) k_B.
\end{equation}
Thus, loosely speaking, the system has gained one degree of freedom 
per field-induced vortex and entanglement length. The result is again 
perfectly consistent with linelike melting.

The discontinuity in magnetization can be computed from the entropy 
jump using the Clausius-Clapeyron relation,
\begin{equation}
\Delta S = - \frac{\Delta B}{4\pi} \frac{d H_m}{dT},
\end{equation}
where $H_m(T)$ is the applied external field at melting. If we use 
the approximation $H_m(T) \approx B(T)$, which again is true for 
$\lambda > a_0$, we can rewrite this as 
\begin{equation}
\frac{\Delta e_c}{\eps_0} = 8\pi \lambda^2 \Delta \rho,
\end{equation}
where $\rho = B/\Phi_0$ is the density of the flux lines. If we 
further assume the latent heat to be independent of the interaction 
range, we obtain a simple expression for the magnetization jump,
\begin{equation}
\Delta B \approx 5.2 \times 10^{-4} \, \Phi_0 /\lambda^2.
\end{equation}
This result explains why the magnetization jump vanishes in the limit 
$\lambda \rightarrow \infty$ and is in good agreement with experimental 
results for YBCO. An advantage with the Bose model is that we can do 
a simulation with finite $\lambda$, allowing the density to 
fluctuate, and test the consistency of the simulation with 
Clausius-Clapeyron directly.\cite{Nordborg-Blatter:PRB98}

It is also interesting to examine what fraction of the latent heat 
comes from the jump in tilting energy of vortices which corresponds to the 
Josephson coupling between the layers in the XY model.  It was shown that 
for the Bose model the jump in the average tilt energy $\Delta 
e_{tilt}$ is exactly half of the latent heat $\Delta e_{c}$, where 
$e_{tilt} = \eps_{l}\langle (d{\bf 
R}/dz)^{2}\rangle/2$.\cite{Nordborg-Blatter:PRB98} This can also be 
considered as an indication of the linelike regime.  To investigate 
this property for the XY model one has to study the relative jump of 
the Josephson energy $\Delta e_{J}/\Delta e_{c}$ with 
$e_{J}=-fJ/(\gamma^{2})\langle \cos [ \phi({\bf n} + {\bf 
d}_z) - \phi({\bf n})] \rangle$.  Simulations shows that for a wide 
range of fields $\Delta e_{J}\approx 0.5 \Delta 
e_{c}$.\cite{Koshelev:PRB97} Moreover, using the fact that the free 
energy must be continuous at melting, the following relation can be 
derived
\begin{equation}
\frac{\Delta e_J}{\Delta e_{c}} = - \frac{\partial \ln T_m}{\partial
\ln \gamma^2},
\end{equation}
which shows that this property is a consequence of the scaling 
relation (\ref{meltlineXY}). 

To conclude, we have compared and discussed numerical result from two 
widely used models for the vortex system, the XY-model and the Bose 
model.  We have shown that the models agree even quantitatively over a 
large part of the phase diagram, where we conclude the melting to be 
linelike.  We have identified a number of universal properties, the 
entanglement length and configurational entropy per entanglement 
length, which we believe are generic to the transition in the linelike 
regime.   

This work was supported by the National Science Foundation Office of 
the Science and Technology Center under contract No.  DMR-91-20000.  
and by the U. S. Department of Energy, BES-Materials Sciences, under 
contract No.  W-31- 109-ENG-38.  The authors gratefully acknowledge 
use of the Argonne High-Performance Computing Research Facility.  The 
HPCRF is funded principally by the U.S. Department of Energy Office of 
Scientific Computing.

\section{Appendix: Fluctuation corrections to helicity moduli in
XY-model}

The helicity moduli $\Upsilon _{x}$ and $\Upsilon _{z}$ describe the 
supercurrent responses to phase rotations.  In the Meissner state they 
are determined by the following thermodynamic averages 
\begin{eqnarray} \Upsilon _{x} &=&J\left\langle V^{\prime \prime 
}\left( \phi ({\bf n}+{\bf d} _{x})-\phi ({\bf n})\right) 
\right\rangle - \frac{J^{2}}{TN}\left\langle \left[ \sum_{{\bf 
n}}V^{\prime }\left( \phi ({\bf n}+{\bf d}_{x})-\phi ({\bf n})\right) 
\right] ^{2}\right\rangle, \label{DefHelx} \\
\Upsilon _{z} &=&\frac{J}{\gamma ^{2}}\left\langle \cos \left( \phi
({\bf n}+
{\bf d}_{z})-\phi ({\bf n})\right) \right\rangle
 - \frac{J^{2}}{TN\gamma ^{4}}
\left\langle \left[ \sum_{{\bf n}}\sin \left( \phi ({\bf n}+{\bf d}
_{z})-\phi ({\bf n})\right) \right] ^{2}\right\rangle,
\label{DefHelz}
\end{eqnarray}
where $N$ is the total number of grid sites, $V^{\prime }=dV(\phi
)/d\phi $ , $V^{\prime \prime }=d^{2}V(\phi )/d\phi ^{2}$, and
$\langle\ldots\rangle$ notates average with the energy functional (1).
At finite temperatures the helicity moduli are suppressed by
fluctuations below their zero temperature values $J$ and $J/\gamma
^{2}$
\begin{eqnarray}
\Upsilon _{x} &\approx &J\left( 1-\frac{V^{(4)}}{2}\left\langle \left(
\phi (
{\bf n}+{\bf d}_{x})-\phi ({\bf n})\right) ^{2}\right\rangle \right)
\label{CorrHelx} \\
\Upsilon _{z} &\approx &\frac{J}{\gamma ^{2}}\left( 1-\frac{1}{2}
\left\langle \left( \phi ({\bf n}+{\bf d}_{z})-\phi ({\bf n})\right)
^{2}\right\rangle \right)   \label{CorrHelz}
\end{eqnarray}
where $V^{(4)}=d^{4}V(\phi )/d\phi ^{4}|_{\phi =0}$ .  Phase 
fluctuations at small temperatures are determined by the Gaussian 
energy functional which is obtained by expansion of Eq.(1)
\begin{equation}
{\cal F}[\phi ({\bf n})] = \frac{J}{2}\sum_{{\bf n}}\left[ \sum_{\alpha
=x,y}\left( \phi ({\bf n}+{\bf d}_{a})-\phi ({\bf n})\right)
^{2} \nonumber + \frac{1}{
\gamma ^{2}}\left( \phi ({\bf n}+d_{z})-\phi ({\bf n})\right)
^{2}\right]  \label{GaussEn}
\end{equation}
Using this functional we calculate
\begin{eqnarray}
\left\langle \left( \phi ({\bf n}+d_{x})-\phi ({\bf n})\right)
^{2}\right\rangle & = & \frac{T}{2J}\left( 1-\frac{1}{2\pi \gamma
^{2}}\left(
\ln 32\gamma ^{2}-1\right) \right)   \label{PhFluctx} \\
\left\langle (\phi ({\bf n}+d_{z})-\phi ({\bf n}))^{2}\right\rangle
& = & \frac{T}{2\pi J}\left( \ln 32\gamma ^{2}-1\right)   \label{PhFluctz}
\end{eqnarray}
Substituting these expressions into Eqs.\ (\ref{CorrHelx}) and
(\ref{CorrHelz}
) we obtain Eqs. (7) and (8) of the paper.


\end{document}